\documentclass[twocolumn,preprintnumbers,prb,aps,amssymb,superscriptaddress]{revtex4}
\usepackage{graphicx}
\usepackage{dcolumn}
\usepackage{bm}
\begin{document}

\newcommand{\Co}{CeCoIn$_5$}
\newcommand{\Rh}{CeRhIn$_5$}
\newcommand{\Ir}{CeIrIn$_5$}
\newcommand{\Tc}{$T_c$}
\newcommand{\ie}{{\it i.e.}}
\newcommand{\eg}{{\it e.g.}}
\newcommand{\etal}{{\it et al.}}


\title{Ambient-pressure bulk superconductivity deep in the magnetic state of CeRhIn$_5$}


\author{Johnpierre~Paglione}
\affiliation{Department of Physics, University of California at San Diego, La Jolla, CA 92093}
\affiliation{Center for Nanophysics and Advanced Materials, Department of Physics, University of Marland, College Park, MD 20742}

\author{P.-C.~Ho}
\affiliation{Department of Physics, University of California at San Diego, La Jolla, CA 92093}
\affiliation{Department of Physics, California State University, Fresno, CA 93740}

\author{M.~B.~Maple}
\affiliation{Department of Physics, University of California at San Diego, La Jolla, CA 92093}

\author{M.~A.~Tanatar}
\altaffiliation[Permanent address: ] {Inst. Surface Chemistry, N.A.S. Ukraine, Kyiv, Ukraine.}
\affiliation{D\'epartement de Physique et RQMP, Universit\'e de Sherbrooke, Sherbrooke, Canada J1K~2R1}

\author{Louis Taillefer}
\affiliation{D\'epartement de Physique et RQMP, Universit\'e de Sherbrooke, Sherbrooke, Canada J1K~2R1}

\author{Y.~Lee}
\affiliation{Department of Earth System Sciences, Yonsei University, Seoul 120749, Korea}

\author{C.~Petrovic}
\affiliation{Department of Physics, Brookhaven National Laboratory, Upton, NY 11973}

\begin{abstract}

Specific heat, magnetic susceptibility and electrical transport measurements were performed at ambient pressure on high-quality single crystal specimens of CeRhIn$_5$ down to ultra-low temperatures. We report signatures of an anomaly observed in all measured quantities consistent with a bulk phase transition to a superconducting state at $T_c=110$~mK. Occurring far below the onset of antiferromagnetism at $T_N=3.8$~K, this transition appears to involve a significant portion of the available low-temperature density of electronic states, exhibiting an entropy change in line with that found in other members of the 115 family of superconductors tuned away from quantum criticality. 

\end{abstract}

\maketitle


With the record-high transition temperature of $T_c=2.3$~K for Ce-based heavy-fermion materials, \cite{Petrovic_Co} superconductivity in \Co\ has been the subject of many studies which have since revealed an exotic pairing state with several intriguing properties including unconventional (nodal) gap symmetry,\cite{Izawa,Aoki} an anomalously large specific heat jump at \Tc,\cite{Petrovic_Co} unconventional behavior in superfluid density,\cite{Ozcan} multiple-size energy gaps,\cite{Rourke} and unpaired quasiparticles in the $T\to 0$ limit.\cite{Tanatar} Furthermore, the application of magnetic field exposes additional anomalies, including a first-order superconductor to normal state transition,\cite{Izawa,Bianchi} and a magnetic quantum critical point which coincides with the upper critical field $H_{c2}$.\cite{Paglione_QCP} \Rh, on the other hand, is a well-characterized\cite{Cornelius,Llobet,Paglione_Rh} antiferromagnet at ambient conditions with a N\'eel temperature $T_N=3.8$~K that is gradually suppressed upon application of pressure\cite{Hegger,Llobet,Knebel04,Park} or Co substitution\cite{Zapf,Jeffries} to reveal a superconducting state that is widely thought\cite{Sarrao} to resemble that found in \Co. This tuning is considered to be strongly tied to the nature of Ce $f$-electron states, since both theoretical \cite{Shishido} and experimental\cite{Shishido,Alver} evidence points to a localized $f$-electron scenario for \Rh\ and a delocalized one for \Co.

Intriguingly, signatures of a low-lying superconducting phase in \Rh\ were observed even at {\it ambient pressure} by Zapf \etal, and more recently by Chen \etal, as a diamagnetic drop in susceptibility measurements near 100~mK.\cite{Zapf,Chen} Lacking any evidence for superconductivity in any of the La- or Y-based (non-$f$-electron) analogs of \Rh, this phase is unlikely to arise from pairing of electronic $d$- and $p$-states, but rather is associated with delocalized $f$-states. However, a non-bulk superconducting phase cannot be ruled out as the cause of the observed low-pressure diamagnetic response. For instance, it has been predicted that superconductivity can be stabilized at the surface of a magnetic material,\cite{Buzdin} where the internal (spontaneous) magnetic field tends toward zero. More important, the third member of the 115 family, \Ir, indeed exhibits a non-bulk superconducting phase with supercurrents causing a resistive transition at 1.2~K, while a thermodynamic signature only appears at a much lower temperature of 0.4~K (Ref.~\onlinecite{Petrovic_Ir}). The likely existence of filamentary \cite{Bianchi_Ir} and/or surface\cite{Fukui} superconducting phases in \Ir\ thus adds to the controversy about ambient-pressure superconductivity in \Rh.

Here we establish the bulk nature of the transition observed deep within the antiferromagnetic state of \Rh\ by presenting the first clear thermodynamic evidence for ambient-pressure superconductivity in this material: deep below the onset of incommensurate magnetic order in \Rh, a significant fraction of the remaining density of electronic states undergoes a bulk superconducting transition at 110~mK. We show that the thermodynamic nature of this transition is quite similar to that found throughout the 115 series, confirming the ubiquitous presence of pairing instabilities in this family of materials even far from criticality.


Large, single-crystal specimens of \Rh\ were grown by the self-flux method,\cite{Petrovic_Co} yielding samples with unprecedentedly low impurity concentrations as evidenced by a residual resistivity $\rho_0 = 37$~n$\Omega$cm and resistivity ratio $\rho(300$~K)/$\rho_0 \sim 1000$ (Ref.~\onlinecite{Paglione_Rh}). Resistivity samples were prepared with dimensions $\sim 4 \times 0.1 \times 0.05$~mm$^3$ and measured with an AC resistance bridge, applying currents parallel to the basal plane of the tetragonal crystal structure. Both AC susceptibility and specific heat measurements were performed on a larger single-phase crystal from the same growth batch, using a conventional low-frequency drive/pickup coil method and a standard semi-adiabatic heat pulse method, respectively. All experiments were performed in a dilution refrigerator. 


\begin{figure}
 \centering
 \includegraphics[totalheight=2.75in]{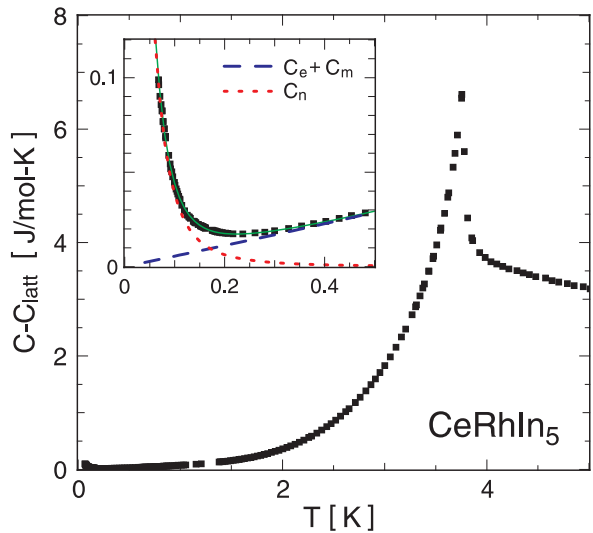}
 \caption{\label{fig:CvsT} Specific heat of \Rh\ with lattice contribution $C_{latt}$ subtracted (see text). Inset shows zoom of low temperature data and fit (solid line) composed of individual electronic plus magnetic ($C_e+C_m$, dashed line) and nuclear ($C_n$, dotted line) contributions as explained in the text.}
\end{figure}

As shown in Fig.~\ref{fig:CvsT}, the specific heat of \Rh\ is characterized by a dramatic transition into the antiferromagnetic state at 3.8~K, as evidenced by the sharp anomaly in $C(T)$. Below $T_N$, $C(T)$ has been previously characterized as composed of a standard electronic contribution $C_e(T)=\gamma T$ with $\gamma=56$~mJ/mol-K$^2$, 
in line with the universal Kadowaki-Woods ratio,\cite{Paglione_Rh} and a magnetic contribution $C_m(T)$, which itself shows evidence of a gapped desity-wave-like state with gap magnitude of $\sim 8$~K.\cite{Cornelius} Here we reproduce the same results, but focus on the  properties far below $T_N$. In this regime, the total specific heat can be described by the sum of several standard contributions:
\begin{equation}
C(T) = C_{latt} +C_e + C_m +  C_n,
\end{equation}
including lattice, electronic, magnetic, and nuclear terms, respectively. Shown in Fig.~\ref{fig:CvsT} is the total measured $C(T)$ less the lattice term, estimated using the measured specific heat of the non-$f$-electron analog YRhIn$_5$ (with $\gamma \ll 56$~mJ/mol-K$^2$). 

Below $\sim 200$~mK a large upturn in $C(T)$ (Fig.~\ref{fig:CvsT} inset) is attributable to a low-lying Schottky anomaly arising from the splitting of the degenerate nuclear energy levels of indium by its large nuclear quadrupole moment. To analyze this low temperature behavior, a fit of $C(T)$ below 2~K was performed by fixing $\gamma=56$~mJ/mol-K$^2$ and fitting both the magnetic and nuclear contributions as follows. To model the magnetic contribution $C_m$, we employed a simple power law (\eg, $C_m(T)=b T^{\beta}$) to characterize $C(T)$ well above the temperatures where $C_n$ is sizeable. This yields a phenomenological exponent $\beta=4.1$, which captures the curvature of $C(T)$ below 2~K while essentially folding in the details of the analysis performed previously.\cite{Cornelius} To model the nuclear component $C_n$, we include the first three terms ($i=1,2,3$) of the series expansion $C_n(T) = \sum_{i}c_iT^{-i}$ to approximate the high-temperature side of the nuclear Schottky peak. This results in an adequate fit to the data at both zero field and applied magnetic fields up to 330~mT (as discussed below) that compares well with other measurements (\eg, $0.290/T^2$~mJ/mol-K, the value of the most significant ($i=2$) term, compares favorably with that of both \Ir\ and \Rh).\cite{Petrovic_Ir,VanHieu}

\begin{figure}
 \centering
 \includegraphics[totalheight=3.3in]{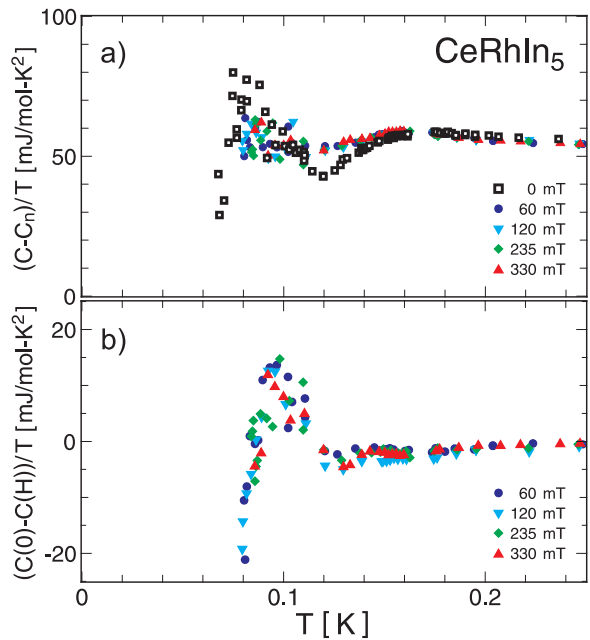}
 \caption{\label{fig:tc} Signature of bulk, ambient-pressure superconducting transition in \Rh, extracted from the total specific heat via (a) subtraction of the nuclear Schottky contribution $C_n$ as obtained from fits to the data (see text) for various magnetic fields, and (b) subtraction of the normal state ($H>H_{c2}$) specific heat $C(H)$ from the total zero-field specific heat $C(0)$.}
\end{figure}

To our surprise, and in contrast to recent reports,\cite{VanHieu}
an examination of the fit residuals reveals an anomaly in the zero field data which is absent in fields of 60~mT and above. As shown in Fig.~\ref{fig:tc}a, subtracting the fitted form of $C_n(T)$ works quite well down to the lowest temperatures for all finite-field data, leaving the same approximately $T$-independent contribution from $\gamma$. However, applying the same procedure to the zero-field data reveals a distinct anomaly in $(C-C_n)/T$ near $\sim 100$~mK, followed by a decrease toward zero. To check that this feature is not an extrinsic result of the fitting procedure, we also performed a subtraction of the finite-field raw data $C(T,H)$ from the zero-field raw data $C(T,0)$, as shown in Fig.~\ref{fig:tc}b. This procedure, which does not involve any fitting procedures aside from simple data interpolation, results in an almost identical jump in the difference $(C(0)-C(H))/T$ as that found in the nuclear fit subtraction (Fig.~\ref{fig:tc}a). 

We therefore conclude that this zero-field anomaly, reminiscent of a BCS-like phase transition that gaps the density of states and sends $C(T)$ toward zero at lower temperatures, provides the first evidence for a bulk, thermodynamic phase transition at $T_c=110$~mK in \Rh. Moreover, this confirms the existence of a robust, ambient-pressure pairing instability in {\it all three members} of the Ce-based 115 family. Before discussing the implications of this, we further verify our observations via susceptibility, resistivity and X-ray experiments.

\begin{figure}
 \centering
 \includegraphics[totalheight=2.4in]{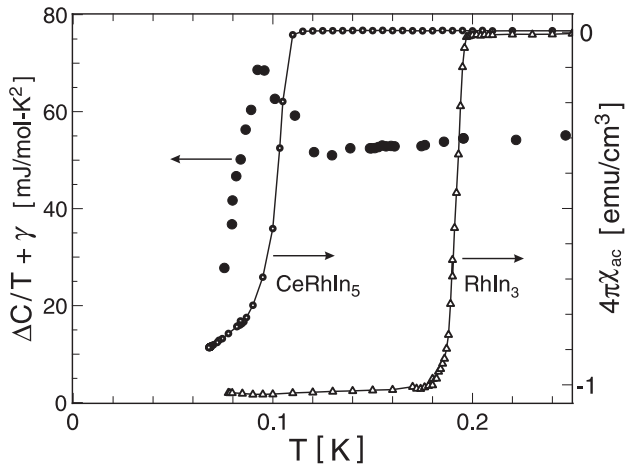}
 \caption{\label{fig:chi} Comparison of extracted specific heat transition (where $\Delta C = C(0) - C(120$~mT) and $\gamma = 56$~mJ/mol-K$^2$); solid circles) and the diamagnetic response of the same 55~mg single-crystal specimen of \Rh\ (open circles), showing the bulk, full-volume nature of the ambient-pressure superconducting transition at $T_c = 110$~mK. The superconducting transition in a similar sized single-crystal sample of RhIn$_3$ is also shown (triangles) for comparison.}
\end{figure}

AC magnetic susceptibility $\chi_{ac}$ measured on the same specimen reveals a diamagnetic response at $T_c$ nearly identical to that reported previously,\cite{Zapf,Chen} confirming the superconducting nature of the transition. As shown in Fig.~\ref{fig:chi}, the drop in $\chi_{ac}$ reaches at least $90\%$ of $-1/4\pi$ by 70~mK and approaches a full 100\% volume fraction as $T\to 0$, as determined by comparing to the response from a similar size specimen of superconducting aluminum. Furthermore, the midpoint temperature of the drop in $\chi_{ac}$ coincides precisely with that of the rise in $C(T)$, confirming the coincidence of $T_c$ in both quantities. 

\begin{figure}
 \centering
 \includegraphics[totalheight=2.4in]{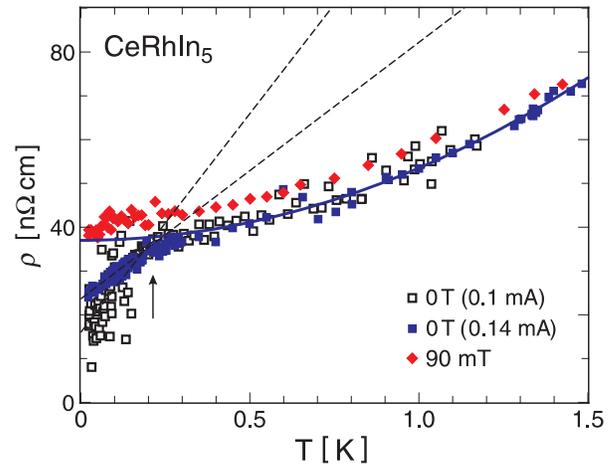}
 \caption{\label{fig:rho} Low temperature resistivity of \Rh\ in zero and applied magnetic fields. Dashed lines are linear fits to zero-field data below the kink associated with the onset (indicated by arrow) of the superconducting transition at \Tc, showing a slight dependence on excitation current. The solid line is a quadratic fit to zero-field data above the kink.}
\end{figure}

Fig.~\ref{fig:rho} presents the low temperature resistivity $\rho(T)$ of \Rh, which exhibits a drop below its saturating value consistent with an onset of superconductivity near $\sim 200$~mK. As shown, the kink in $\rho(T)$ is discernible when using an excitation current of $100~\mu$A, which is at the threshold of our ability to perform this measurement.\cite{nonzero} This anomaly is consistent with the resistive onset of a superconducting transition centered at $T_c=110$~mK, assuming a resistive transition width of $\simeq 100$~mK similar to that typically found in other 115 materials.\cite{Petrovic_Co,Petrovic_Ir}  As shown in Fig.~\ref{fig:rho}, the application of a larger current ($140~\mu$A) improves signal to noise but also appears to supress the transition, as does a 90~mT field.

Finally, to rule out the possibility of impurity phase contamination, we have grown a high purity ($\rho(300$~K)/$\rho_0>100$) single-crystal specimen of RhIn$_3$, a tetragonal compound\cite{Pottgen} which we suspected may appear in the \Rh\ growth process. As shown in Fig.~\ref{fig:chi}, this material indeed exhibits a diamagnetic full-volume transition at $T_c=190$~mK, providing the first evidence of superconductivity in this material.\cite{Matthias} However, because there is no visible anomaly in any measured quantities of \Rh\ at this temperature, RhIn$_3$ is unlikely to be involved in the observation of superconductivity in our samples of \Rh.  Nevertheless, to completely rule out this possibility, high-resolution synchotron powder X-ray measurements on a piece of the same crystal of \Rh\ used for $C(T)$ and $\chi(T)$ measurements confirm no detectable trace of RhIn$_3$. 

Together, this evidence conclusively rules out the possibility of filiamentary, surface or impurity superconductivity in \Rh, confirming the bulk nature of superconductivity in \Rh\ previously suggested to compete with magnetism.\cite{Chen} Indeed, pairing in \Rh\ appears to arise from the quasi-particle density of states left over from the magnetic ordering mechanism. As shown in Fig.~\ref{fig:chi}, the jump in specific heat associated with \Tc\ is approximately half the size of $\gamma$ itself, with a value $\Delta C/C(T_c) \simeq 0.5$. In a superconductor, this jump is a result of the sudden loss of entropy due to pairing and can vary widely from the weak-coupling BCS expectation of 1.43 for $s$-wave superconductors, being either enhanced due to strong coupling or reduced by the presence of gap anisotropy. Although the ambient-pressure transition in \Rh\ lies deep within an apparent Fermi liquid state (as evidenced by $T^2$ scattering below $\sim 2$~K),\cite{Paglione_Rh} the value $\Delta C/C(T_c) \simeq 0.5$ is much smaller than the weak-coupling BCS expectation for $s$-wave symmetry and is more consistent with a $d$-wave order parameter in the weak-coupling limit.\cite{Carbotte}

Interestingly, while being almost an order of magnitude smaller than the anomalously large value of 4.5 observed in \Co,\cite{Petrovic_Co} the jump in \Rh\ is actually quite comparable to that measured in the same material under applied pressures.\cite{Knebel04} In fact, it is exactly the same as that found at 1.90~GPa -- the pressure where $T_N$ is suppressed to the same value as \Tc\ ($\sim 2$~K), and above which the manetic transition is no longer detectable in zero field.\cite{Park} This provides evidence that the superconducting state observed at zero pressure does not significantly change its nature upon pressure increase, at least up to the point where $T_N \simeq T_c$ (assuming no significant change in the coupling strength with pressure). Rather, the density of electronic states available for pairing gradually increases, along with \Tc, in balance with a decreasing $T_N$. This is indicative of a competition for Fermi surface betweem the incommensurate-antiferromagnetic and superconducting ground states, as recently advocated both experimentally\cite{Chen} and theoretically\cite{Alvarez} for \Rh, and proposed over two decades ago for the heavy-fermion superconductor URu$_2$Si$_2$, in which the Fermi surface is partially gapped.\cite{Maple}

However, a qualitative change in this scenario appears to occur in \Rh\ once magnetism is suppressed.
Rather than remaining constant upon further pressure increase, $\Delta C/C(T_c)$ undergoes a substantial threefold growth -- reaching a maximum of $\sim 1.45$ at 2.55~PGa (coincident with the maximum \Tc) before decreasing back toward $\sim 0.5$ at much higher pressures (\ie\ near 3.38~GPa, where $T_c \simeq 2.0$~K).\cite{Knebel04} In other words, the superconducting state of \Rh\ near its pressure-tuned QCP qualitatively differs from that found under both low- and high-pressure conditions, while being strikingly similar to that of \Co\ at ambient pressure. Interestingly, the third member of this family, \Ir, also has a small $\Delta C/C(T_c) = 0.76$ at ambient pressure,\cite{Petrovic_Ir} provoking the idea that its pairing state differs from those in the critical pressure region yet perhaps shares similarities with \Rh\ at either low or extremely high pressures. Recent thermal conductivity measurements have in fact shown evidence for different (bulk) superconducting gap symmetries in \Co\ and \Ir.\cite{Shakeripour} However, the jump in \Ir\ differs from that in both its Rh- and Co-based counterparts, exhibiting a minimal change with increasing pressure\cite{Borth} that suggests (at the very least) that the pairing state in \Ir\ is much less sensitive to pressure. In any case, the evidence for multiple, distinct superconducting phases in this family is further supported by the observation of separated ``domes'' of superconductivity as a function of pressure in CeRh$_{1-x}$Ir$_x$In$_5$.\cite{Nicklas} It remains to be determined whether the competition between magnetic and superconducting states for electronic density of states in this family is due to a gapped nature of magnetism, as recently proposed theretically,\cite{Alvarez} or a different phase space distribution for each coupling mechanism.


In conclusion, the first evidence of a bulk-phase superconducting state in \Rh\ at ambient pressure is provided by the observation of a phase transition in specific heat that occurs simultaneously with both a diamagnetic transition in magnetic susceptibility and a drop in electrical resistivity. With evidence for an ambient-pressure superconducting ground state in all three members of the Ce-based 115 materials, critical comparisons of each may play a vital role to elucidating the origins of both unconventional superconducting and normal state characteristics throughout this family.


The authors thankfully acknowledge K.~Behnia and V.S.~Zapf.
Crystal growth and characterization was carried out at the Brookhaven National Laboratory, operated for the U.S. DOE by Brookhaven Science Associates DE-Ac02-98CH10886. 
Low temperature studies were funded by NSF (DMR-03-35173) and NSERC of Canada.
J.P. acknowledges support from a NSERC Canada postdoctoral fellowship.



\begin{references}
\bibitem{nonzero} The measurement of a complete zero-resistance state in \Rh\ remains elusive due its extremely high conductivity (\ie, $<0.1~{\rm m}\Omega$ sample resistance).
\end{references}
\end{document}